\documentclass[aps,amsmath,amssymb,preprint,floats]{revtex4}
\usepackage{amsmath}
\usepackage{graphicx}
\usepackage{dcolumn}
\usepackage{bm}
\usepackage[usenames]{color}

\begin{document}
\newcommand{\be}{\begin{equation}}
\newcommand{\ee}{\end{equation}}
\newcommand{\ben}{\begin{eqnarray}}
\newcommand{\een}{\end{eqnarray}}
\newcommand{\nn}{\nonumber \\}
\newcommand{\ii}{\'{\i}}
\newcommand{\pp}{\prime}
\newcommand{\expq}{e_q}
\newcommand{\lnq}{\ln_q}
\newcommand{\quno}{q-1}
\newcommand{\qunoinv}{\frac{1}{q-1}}
\newcommand{\tr}{{\mathrm{Tr}}}
\newcommand{\nd}{\noindent}

\title{Aspects of quantum phase transitions}

\author{M. K. G.
Kruse$^{1}, $H. G. Miller$^{1}$\footnote{hmiller@maple.up.ac.za},
A. Plastino$^{2}$\footnote{plastino@fisica.unlp.edu.ar}, A. R.
Plastino$^{1,\,3,\,4}$\footnote{angel.plastino@up.ac.za}, }
\affiliation{
$^1$Department of Physics, University of Pretoria - 0002 Pretoria,
South Africa \\$^2$National University La Plata
(UNLP)\\
  IFLP-CCT-Conicet, C.C. 727, 1900 La Plata, Argentina\\
$^3$Instituto Carlos I de F\'{\i}sica Te\'orica y Computacional,
Universidad de Granada, Granada, Spain\\$^4$CREG-UNLP-Conicet,
Argentina}

\date{\today}
\begin{abstract}
  A unified description of i) classical phase transitions and their remnants in
finite systems and ii) quantum phase transitions is 
presented. The ensuing discussion relies on the interplay between,
on the one hand, the thermodynamic concepts of temperature and
specific heat and on the other,  the quantal ones  of
coupling strengths in the Hamiltonian. Our considerations are
illustrated in
 an exactly solvable model of
Plastino and Moszkowski [Il Nuovo Cimento {\bf 47}, 470 (1978)].
\end{abstract}


\maketitle

\section{Introduction}
In infinite as well as in  finite systems a type of phase
transition, often referred to as a quantum phase transition (qpt),
may occur at T=0. Such quantum phase transitions differ from
classical phase transitions, which can happen only in an infinite
systems at T$\neq$0, and generally signal a change in the
correlations present in the ground state of the system. For an
infinite system described by a Hamiltonian, $H(\lambda)= H_0+\lambda
H_1$, which varies as a function of the coupling constant $\lambda,$
the presence of a qpt can easily be understood in the following
manner\cite{S99}. Generally the ground state energy  is an analytic
and monotonic function of $\lambda$. However, if $[H_0,H_1]= 0$,
level crossing may come about and the ground state energy is no
longer analytic nor monotonic. Although there are other valid
mathematical reasons that lead to the loss of analyticity\cite{S99},
the above simple explanation will suffice for our purposes and
provides a simple means for defining a qpt in an infinite system.  At
some critical value of the coupling constant, $\lambda_c$, a new
ground state comes to pass. For $T > 0$ two possibilities exist:
$\lambda_c$ is an isolated point and the rest the phase diagram is
analytic (wrt $\lambda$) or a classical phase transition may occur.
In the latter case, for example, for a second order phase
transition, the free energy is no longer an analytical function of
$\lambda$. As one varies $\lambda$ a line of singularities occurs at
different temperatures which terminates at T=0 at $\lambda_c$. This
provides a simple means of determining $\lambda_c$, the critical
value at which a qpt occurs in an infinite system.

In finite systems a qpt can take place, but strictly speaking
classical phase transitions can not, since at finite temperatures
the partition function and all related quantities are analytic.  At
best only the remnant  of a classical phase transition may
exist\cite{DM87}. Furthermore, thermal fluctuations about
equilibrium values are large\cite{AZ84} particularly in the region
where this remnant occurs. For example,  studies of their effect  on
an order parameter have concluded that, in atomic nuclei, the
super-conducting to normal phase transition is washed
out\cite{ERIM85,G84}.  However, in spite of these problems a phase
diagram has been constructed from the remnants in an exactly
solvable model\cite{DM87} by studying the specific heat, C.

Clearly information about classical phase transitions or their remnants is
contained in C.  As $T \rightarrow 0$, however, $C\rightarrow 0$. In spite of
this we will show that it is possible to extract information about qpts by
studying C in the limit when $T \rightarrow 0$.  Only some
elementary concepts from Information Theory are required.

\section{Formalism}
\subsection{General considerations}

Consider a system whose dynamics is described (at T=0) by the
following Hamiltonian operator
\begin{equation} \label{i1}
 \hat{H}=\hat{H}_0 + \lambda \hat{H}_1
\end{equation}
where $[\hat{H}_0,\hat{H}_1] =0$. At finite temperatures, the
Maximum Entropy Principle of Jaynes\cite{J57,J57a} can be used to
determine the appropriate statistical operator, $\hat{\rho}$ in the
following manner. Maximizing  the entropy,
$S(\hat{\rho})=Tr[\hat{\rho}\log\hat{\rho}]$,
\begin{equation} \label{i2}
\delta_\rho S(\hat{\rho})=0
\end{equation}
subject to the constraints
\begin{equation} \label{i3}
 <\hat{H}>=Tr[\hat{\rho}\hat{H}]= \mathcal{E}
\end{equation}
and
\begin{equation} \label{i4}
 Tr[\hat{\rho}]=1
\end{equation}
yields
\begin{equation} \label{i5}
 \hat{\rho}=\frac{\exp^{-\beta \hat{H}}}{\mathcal{Z}}
\end{equation}
where
\begin{equation} \label{i6}
 \mathcal{Z}=Tr[e^{-\beta\hat{H}}].
\end{equation}
Generally, in statistical mechanics the coupling constant $\lambda$
is taken to be a constant and equation({\ref{i2}) is used to
determine the Lagrange multiplier $\beta$.  However, in the case of
a qpt, $\lambda$  is no longer
 constant and a functional relation between  $\beta$ and
$\lambda$ may be obtained, using  equation({\ref{i3}).

The specific heat is given by
\begin{eqnarray}
  C &=& -\beta^2(\frac{\partial <\hat{H}>}{\partial\beta})_\lambda \label{i9}\\
     &=& -\beta^2\frac{\partial\lambda}{\partial \beta}(\frac{\partial
<\hat{H}>}{\partial\lambda})_\beta \label{i9bis}
\end{eqnarray}
and a necessary and sufficient condition for it to vanish at $T = 0$ is
\begin{equation} 
 (\frac{\partial}{\partial \beta}<\hat{H}>)_\lambda=0 \label{cb}
\end{equation}
or equivalently
\begin{equation} 
 \frac{\partial\lambda}{\partial \beta}(\frac{\partial}{\partial
\lambda}<\hat{H}>)_\beta=0 .\label{cl}
\end{equation}
Clearly   $\lambda_{c}$, the critical value of the coupling constant
at T=0, can be determined from equation(\ref{cb}) which clearly indicates that
information about the qpt is contained in the specific heat. On the other hand C
will
vanish in this limit   if
\begin{equation}
\frac{\partial\lambda}{\partial \beta}=0 \label{fl}
\end{equation}
for all values of $\lambda$ (see equation (\ref{cl})) . We therefore suggest
(and will show) that information about a qpt should therefore be contained in
the factor $(\frac{\partial <\hat{H}>}{\partial\lambda})_{T=0}$.
Note, however, that
\begin{equation} \label{i10}
 (\frac{\partial <\hat{H}>}{\partial\lambda})_{T=0}=\frac{\partial
E_{gs}}{\partial\lambda} .
 \end{equation}
since only the ground state is populated at that temperature. If,
indeed as has already been pointed out, a qpt  occurs at a level
crossing then two possibilities exist: 1) a discontinuous
derivative
\begin{equation} \label{G}
G(\lambda)=(\frac{\partial
E_{gs}}{\partial\lambda})_{\beta=\infty}(\lambda), \ee if
$\frac{\partial E_{gs}}{\partial\lambda}$  does not change sign
when passing through $\lambda_c$,  or 2) a null derivative,  if
$\frac{\partial E_{gs}}{\partial\lambda}$  does change sign when
passing through $\lambda_c$.

 Hence, one has a very nice
unified means of identifying  both phase transitions and quantum
phase transitions.  Furthermore, it is not necessary to begin at
finite temperatures to find where a qpt takes place .

For finite systems at finite temperatures
 ($T\neq 0$), C
is analytic and structures in
$\frac{\partial <E>}{\partial\beta}$ should
be indicative of the remnant of a phase
transition. Eq.(\ref{cb})
allows one to correctly determine the position of the qpt.
Alternatively
$ \frac{\partial E_{gs}}{\partial\lambda}$    can be used in the manner
outlined above to determine the position of a qpt
 (see illustrative graphs in the examples discussed
below). These two procedures
 should be equivalent.

\section{The Plastino-Moszkowski model}

This an exactly solvable N-body, SU(2) two-level model
\cite{PM78}. Each level can accommodate $N$ particles,
i.e., is $N-$fold degenerate.
  There are two levels separated by an energy gap $\mathbb{E}$ occupied by $N$
  particles. In the model  the angular momentum-like
  operators $J^2, J_x, J_y, J_z$, with $J(J+1)=N(N+2)/4$ are used.
  The
Hamiltonian to be here employed reads
\begin{equation} \label{i16}
 H = \mathbb{E} J_z - \xi [J^2 - J_z^2 -N/2],
\end{equation}
and its eigenstates  are usually referred to as Dicke-states
\cite{D54}. For convenience we set $\mathbb{E}=1$ and

\begin{equation} \label{i17}
J_z=
(1/2)\sum_{i=1}^{N}\,\sum_{\sigma=1}^{2}\,a^+_{i,\sigma}\,a_{i,\sigma},
\end{equation}
with corresponding expressions for $J_x,\,J_y$.
This is a simple yet nontrivial case of the Lipkin model
\cite{LMG65}. For now, we will only discuss the model in the
zero-temperature regime. The operators appearing in the model
Hamiltonian form a commuting set of observables and are thus
simultaneously diagonalizable.

The ground state of the unperturbed system ($\xi=0$ and at $T=0$)
is
$\left|J,J_z\right\rangle=\left|\frac{N}{2},-\frac{N}{2}\right\rangle$
with the eigenenergy $E_0=-\frac{1}{2}N$. When the interaction is
turned on ($\xi\neq0$) and gradually becomes stronger, the ground
state energy will in general be different from the unperturbed
system for some critical value of $\xi$ that we will call
$\lambda_c$. This sudden change of the ground state energy
signifies a quantum phase transition. It should be noted that for
a given value of $N$, there could be more than one critical point.
The critical values of the $n$th transition, i.e.,  $\lambda_c$ at
that  point, can be found from equation \ref{Hcrits} below,
provided that $\lambda_c > 0$ and $\lambda_c \neq\infty$.

\begin{equation}
\lambda_{c,n}=\frac{1}{N-(2n-1)}. \label{Hcrits}
\end{equation}

\subsection{The $N=2$ problem}

We consider first this simple case, since it can be solved
analytically. Here the J = 1-multiplet for two particles is \{$J_z
= -1 ; 0 ; +1$\}. If we label with the letter {\it i} the three
pertinent $J_z-$ eigenstates one has \{$H_{ii} = -1 ;\ -\xi ;\
+1$\} and \{$\epsilon_i\} = -1 ;\ -\xi ;\ 1$\}, respectively and
\begin{equation}
Z = e^{-\beta} + e^{\beta\xi} + e^{\beta}
\end{equation}
with
\begin{equation}
\frac{\partial Z}{\partial\beta} = e^\beta +\xi e^{\beta\xi} - e^{-\beta}
\end{equation}
Moreover,
\begin{equation}
 Tr[\rho H]=< E >= Z ^{-1} [-e^\beta - \xi e^{\beta\xi} + e^{-\beta} ],
\end{equation}
and
\begin{equation}
Z\frac{\partial< E >}{\partial\beta}= -[e^\beta +\xi^2 e^{\beta\xi}
+e^{-\beta}]-Z^{-1} [-e^\beta-\xi e^{\beta\xi} + e^{-\beta} ][
\frac{\partial Z}{\partial\beta} ].
\end{equation}
Setting $Z\frac{\partial<E>}{\partial\beta}=0$ yields
\begin{equation}
 0=-Z[e^\beta +\xi^2 e^{\beta\xi} + e^{-\beta}]+[e^\beta+\xi e^{\beta\xi} -
e^{-\beta} ][ e^\beta +\xi e^{\beta\xi} - e^{-\beta}],
\end{equation}
i.e.,
\begin{equation}
[2 \cosh \beta + e^{\beta\xi} ][2 \cosh \beta + \xi^2 e^{\beta\xi}
] = [2 \sinh \beta + \xi e^{\beta\xi} ]^2 \label{ceq}
\end{equation}
which is the desired function linking $\xi$ with $\beta$.
Consider now the T = 0 limit, in which  $\beta \rightarrow
\infty,\ \cosh \beta \rightarrow e^\beta ,\ \sinh \beta
\rightarrow e^\beta $. In this limit (\ref{ceq}) becomes \be
(2e^\beta + e^{\beta\xi} )(2e^\beta + \xi^2 e^{\beta\xi} ) =
(2e^\beta + \xi e^{\beta\xi} )^2 = 4e^{2\beta} + 4\xi e^\beta
e^{\beta\xi} + \xi^2 e^{2\beta\xi}, \ee entailing 
 \be          (\xi - 1)^2 = 0  \,\, \Rightarrow
                                \xi = 1,
\ee yielding the exact $\xi$-value at which the qpt takes place,
as demonstrated in \cite{PM78}.

Note, however,  one could alternatively
start with
\begin{equation}
G(\xi)=\frac{\partial< E >}{\partial\xi}= \frac{1}{Z^2}[-(1+\beta)
e^{\beta\xi}Z-(-e^\beta -\xi e^{\beta\xi} +e^{-\beta})\beta e^{\beta\xi}.
].\label{xc}
\end{equation}
Requiring
\begin{eqnarray}
G(\xi)=0
\end{eqnarray}
one obtains in the limit $T\rightarrow 0$ 
\begin{equation}
 e^{\beta}(1-\xi)=0
\end{equation}
or
\begin{equation}
 \xi=1!
\end{equation}
which is the exact $\xi$-value at which the qpt takes place. (Note that at
$\xi=1 \ G(\xi)$ changes sign.)  In
accordance with previous considerations revolving around Eq.
(\ref{G}), it is clear that,  at $T=0$, the function $G$
above suffers a brutal discontinuity at $\xi=\xi_c=1$, since it is
"infinite" everywhere except there, where it vanishes.

\section{Numerical results}
Let us now discuss the numerical results for the model given in
equation \ref{i16}. In this section we set $\xi \equiv \lambda$. We
will consider the case of four and eight particles, respectively.
The Hamiltonian is constructed by employing the standard angular
momentum matrices in the appropriate $J$-multiplet and is then
diagonalized. The resulting $2J+1$ eigenenergies are in general a
function of the coupling constant $\lambda$. This dependence on the
coupling constant ultimately allows for a level crossing to take
place at a critical value of $\lambda_c$. In figure
\ref{fig:n4energy} we have shown the subset of eigenenergies that
lead to two level crossings (qpt's) in the $N=4$ particle case. Note that the
slope of the ground state energy does not change sign.

\begin{figure}
    \centering
    \includegraphics{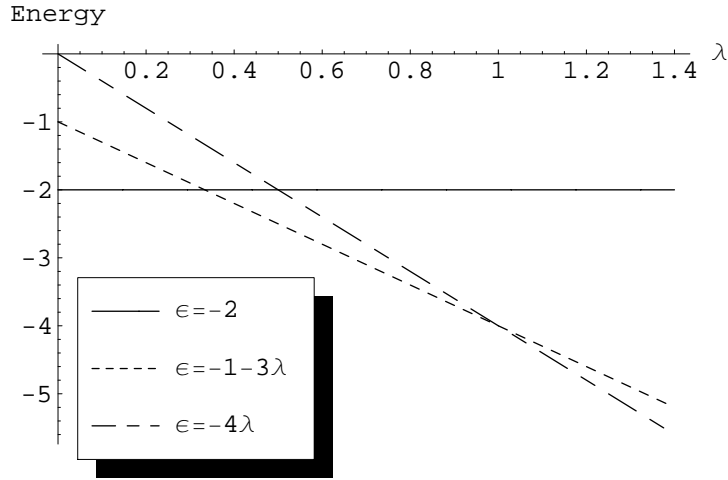}
    \caption{The lowest three eigenenergies of the
    $N=4$ case have been plotted as a function of the
    coupling constant $\lambda$. There are
    level crossings at
    $\lambda_{c,1}=\frac{1}{3}$ and at $\lambda_{c,2}=1$,
    which is in agreement with equation \ref{Hcrits}.
    The eigenenergies of the full system are
    $\epsilon= {\pm2,\pm1-3\lambda,-4\lambda}$.
    The solid, short-dashed and long-short-dashed line correspond to the
eigenenergies
     $\epsilon=-2$, $\epsilon=-1-3\lambda$ and $\epsilon=-4\lambda$,
respectively.}
    \label{fig:n4energy}
\end{figure}

We then construct the canonical partition function $\mathcal{Z}$
from the full set of eigenvalues. One can now
determine the specific heat as given by the two equations
\ref{i9}-\ref{i9bis}.

\subsection{The analogous "specific heat" $C^*_\beta$}
Once the partition function has been constructed from the
eigenvalues of the $N$-particle Hamiltonian, we are able to form the
expectation value of the energy as given by the familiar canonical
ensemble relation below.

\begin{equation}
\mathcal{E}=-\frac{\partial}{\partial\beta}\ln\mathcal{Z}
\end{equation}

The quantity that will be used to map out the phase diagram of the
model, which we will call $C^*_{\beta,\lambda}$, is given by the
derivative of $\mathcal{E}$, with respect to either $\beta$ or
$\lambda$. In this section we will focus our attention on the former
case.

\begin{equation}
C^*_\beta=\frac{\partial}{\partial\beta}\mathcal{E(\beta,\lambda)}
\end{equation}

A plot of $C^*_\beta$ is given in figure \ref{fig:n4cb110} for a
fixed value of $\beta=110$. The value of $\beta$ was an arbitrary
choice, in order to demonstrate the following point. At finite
temperatures, that is when $\beta\neq\infty$, the peaks that are
found in figure \ref{fig:n4cb110} are a signature of a phase
transition taking place. They are smoothed out due to finite
temperature effects. As the temperature is lowered ($\beta$
increases), the peaks move together and become smaller in size. This
is shown in figure \ref{fig:n4peak1}. When $\beta\rightarrow\infty$,
the peaks around each critical point coalesce into a single point,
namely $\lambda_c$. This is exactly what one would expect at zero
temperature; the phase transition takes place where the
eigenen-ergies become degenerate.

\begin{figure}
    \centering
        \includegraphics{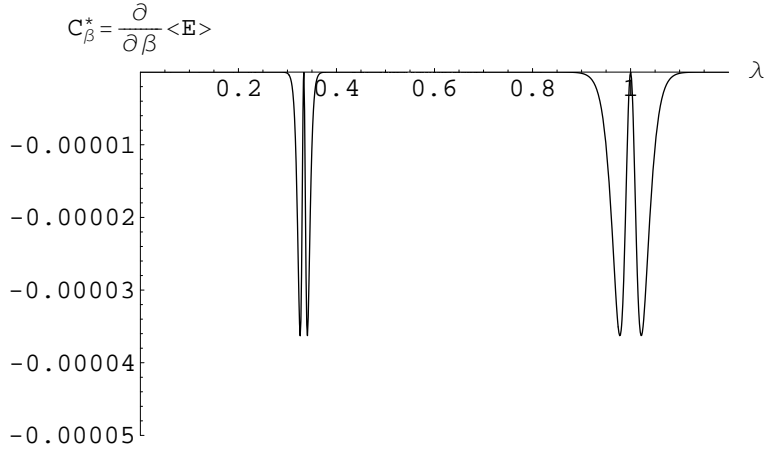}
    \caption{The quantity $C^*_\beta$ has been plotted as a function
    of the coupling constant $\lambda$, for the $N=4$ particle case, with a
fixed value of
    $\beta=110$ (see text for a discussion on this point).
    There are two peaks present in the plot, centered around
    the two critical points $\lambda_c$ of the system.
    The peaks coalesce into a single point centered
    at $\lambda_c$ as $\beta\rightarrow\infty$ (see figure \ref{fig:n4peak1}).
    Everywhere else $C^*_\beta=0$, in agreement with equation \ref{cb}.}
    \label{fig:n4cb110}
\end{figure}

\begin{figure}
    \centering
        \includegraphics{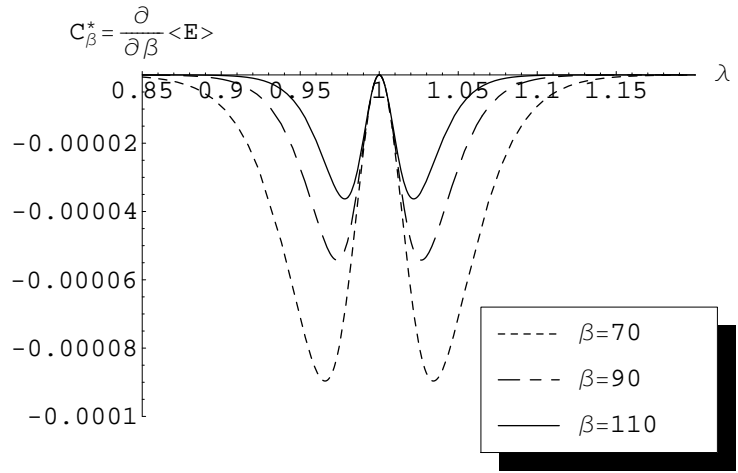}
    \caption{The temperature dependence of $C^*_\beta$
    in the region of $\lambda_c=1$ for the $N=4$ particle
    case has been plotted. The short-dashed, long-short-dashed and solid
    peaks correspond to $\beta=70,90,110$ respectively. One
     can clearly see that as the temperature
     is lowered ($\beta\rightarrow\infty$), that the peaks become
     smaller in size and narrower in width.
     In the zero-temperature limit, these peaks would coalesce
     into a single point situated exactly at the location of
     the quantum phase transition. }
    \label{fig:n4peak1}
\end{figure}

\subsection{The analogous "`specific heat"' $C^*_\lambda$}
The above investigation of the quantity $C^*_\beta$ is one way to
characterize the quantum phase transitions. It is also possible to
investigate the qpt's from another viewpoint. In this section we
will consider the quantity
$C^*_\lambda=\frac{\partial}{\partial\lambda}\mathcal{E(\beta,\lambda)}$.

In figure \ref{fig:n4cvlb} we have plotted the dependence of
$C^*_\lambda$ on $\beta$ for various values of $\lambda$. It can be
seen that if the coupling constant is set in a range corresponding
to one particular value of the ground state eigenenergy, that at
low temperatures $C^*_\lambda$ tends to the value of the slope of
the given eigenenergy. For example, when $0\leq\lambda<\frac{1}{3}$,
$C^*_\lambda\rightarrow0$ as $\beta$ becomes large. For that range
of the coupling constant, the corresponding ground-state eigenvalue
is $\epsilon=-2$, which of course has a slope of zero. Similarly for
$\frac{1}{3}<\lambda<1$, $C^*_\lambda\rightarrow-3$, which
corresponds to the slope of the ground state eigenvalue
$\epsilon=-1-3\lambda$. At the critical values $\lambda_c$,
$C^*_\lambda$ takes on the average value of the slope of the two
degenerate eigenenergies involved. In figure \ref{fig:n4cvll}, we
have plotted the zero temperature limit of $C^*_\lambda$ as a
function of $\lambda$. There are two discontinuities in the figure,
corresponding to the values of $\lambda_c$ where the qpt takes
place. The horizontal  lines in the figure correspond to the slope
of the current ground state eigenvalue.

\begin{figure}
    \centering
        \includegraphics{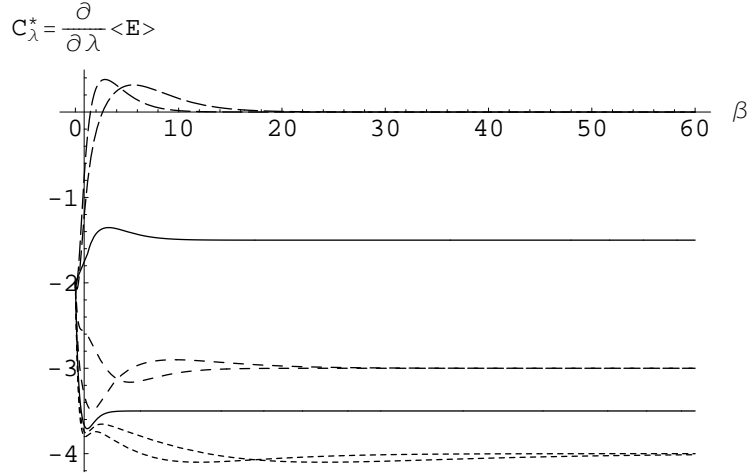}
    \caption{ $C^*_\lambda$ as a function of $\beta$
    for various values of $\lambda$ for the $N=4$ particle case. The long-dashed
curves (top two curves)
    correspond to $\lambda=0.1,0.2$;
    the medium-dashed curves correspond to $\lambda=0.5,0.75$ (in between the
two solid curves);
    the short-dashed curves (lowest two curves) correspond to $\lambda=1.1,1.2$.
    The solid curves correspond to the the critical values of
    $\lambda_{c,n}=\frac{1}{3},1$. Curves that have same dashing style
    correspond to the same ground state eigenvalue and in the
    zero-temperature limit tend to the slope of that corresponding eigenvalue.
    At the critical points, $C^*_\lambda$ picks out the average value
    of the two slopes from the relevant degenerate eigenvalues.}
    \label{fig:n4cvlb}
\end{figure}

\begin{figure}
    \centering
        \includegraphics{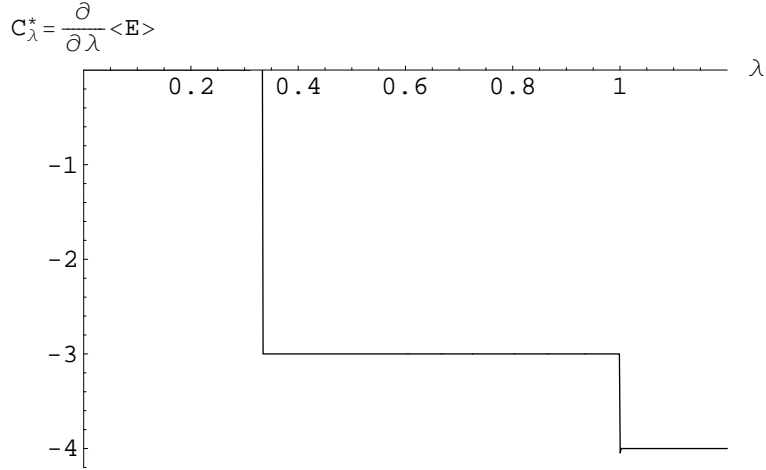}
    \caption{ $C^*_\lambda$ as a function of $\lambda$ in the
    zero-temperature limit for the $N=4$ particle case. The horizontal segments
of the plot
    correspond to the derivatives (with respect to $\lambda$) of
    the ground state eigenvalue for that particular range of $\lambda$.
    For example, $\frac{1}{3}<\lambda<1$, $C^*_\lambda=-3$,
    which corresponds to the slope of the ground state
    eigenvalue $\epsilon=-1-3\lambda$. The discontinuities take place
    at the critical values of the system, viz $\lambda_{c,n}=\frac{1}{3},1$,
     respectively. At the qpt, the value of $C^*_\lambda$ is the average
     value of the two slopes of the relevant degenerate eigenvalues. }
    \label{fig:n4cvll}
\end{figure}

\subsection{The Plastino-Moszkowski model for $N=8$ particles.}

It is also of interest to see if the above methodology works for a
larger system. In this case the
slope of the ground state energy  as a function of the coupling constant
does not change sign. We will briefly summarize the results
when the model
has $N=8$ particles present. Using equation \ref{Hcrits}, we
determine that the critical coupling constants are the following
values: $\lambda_{c,n}=\frac{1}{7},\frac{1}{5},\frac{1}{3},1$. For
completeness, the 9 eigenvalues of the system are
$\epsilon=\pm4,\pm1-15\lambda,\pm2-12\lambda,\pm3-7\lambda,-16\lambda$.
The quantity $C^*_\beta$ is shown in figure \ref{fig:n8cb110} and is
seen to correctly identify where the quantum phase transitions
occur. In figure \ref{fig:n8cvll} we have plotted $C^*_\lambda$ in
the zero-temperature limit as a function of $\lambda$. As in the
$N=4$-particle case, the discontinuous jumps seen in the plot
correspond to a quantum phase transition taking place.

\begin{figure}
    \centering
        \includegraphics{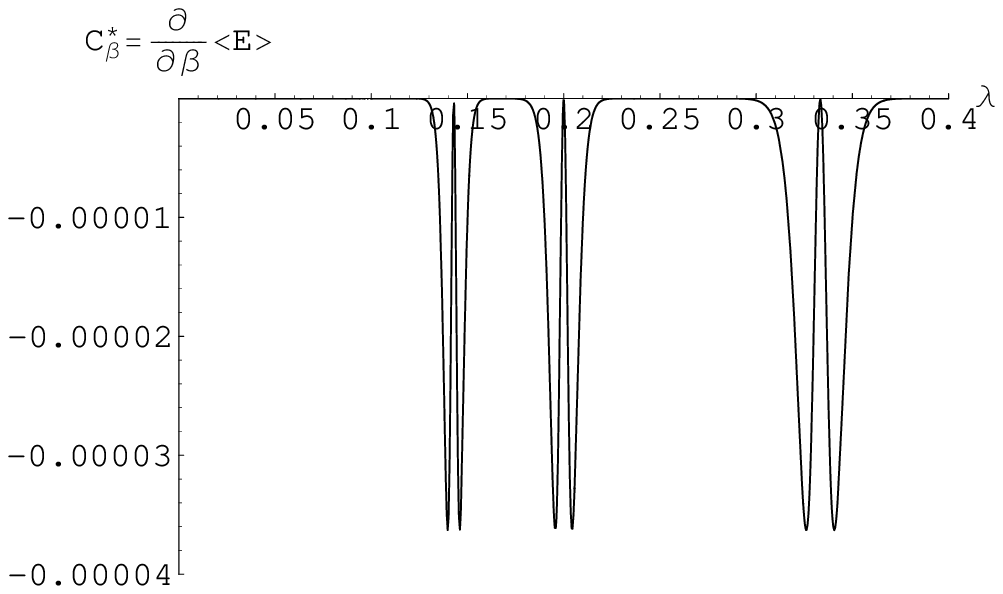}
    \caption{$C^*_\beta$ has been plotted for $\beta=110$ (for illustration)
    in the $N=8$ particle case. The peaks are centered around the critical
    coupling constants $\lambda_{c,n}=\frac{1}{7},\frac{1}{5},\frac{1}{3}$.
    Recall that in the zero-temperature limit the peaks coalesce into a single
    point located at the critical points, as has been already shown for the
    $N=4$ particle case (see figure \ref{fig:n4peak1}). Note that we have only
    plotted the first 3 critical points to make the plot clearer;
    the peak located at $\lambda_c=1$ is not shown.}
    \label{fig:n8cb110}
\end{figure}

\begin{figure}
    \centering
        \includegraphics{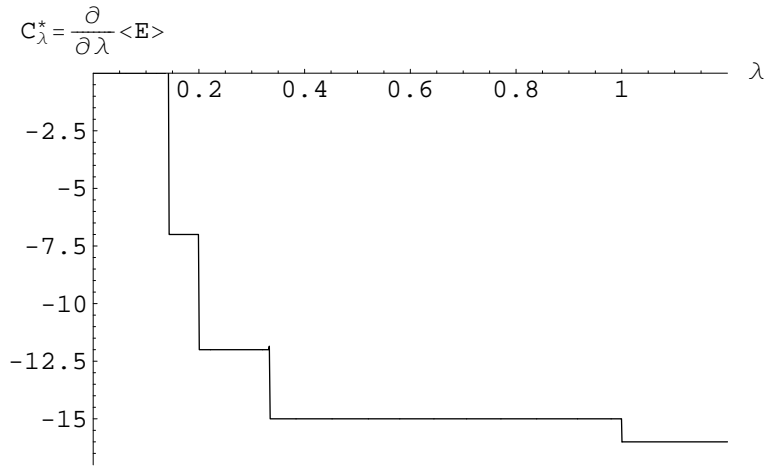}
            \caption{$C^*_\lambda$ has been plotted in the zero-temperature
limit for
            the $N=8$ particle case. The horizontal segments correspond to the
            derivative (with respect to $\lambda$) of the relevant ground state
            of the system for that particular range of $\lambda$.
            The discontinuities take place at the critical values of
            the coupling constant, viz
            $\lambda_{c,n}=\frac{1}{7},\frac{1}{5},\frac{1}{5},1$.
            At the qpt, the value of $C^*_\lambda$
            is the average value of the two slopes of the relevant degenerate
eigenvalues.}
    \label{fig:n8cvll}
\end{figure}

\section{Conclusions}
 We have here shown that classical phase transitions and quantum
phase transitions can be described in a unified fashion. Our treatment has
relied
 heavily  on the specific heat and is also valid for finite systems where only
the remnant of a classical phase transition exists. The pertinent
considerations were  illustrated in
 an exactly solvable model of
Plastino and Moszkowski. In particular we have shown that
information about qpt's can be obtained from the quantity
$\frac{\partial E_{gs}}{\partial\lambda}$ and that this equivalent
to looking at the zero temperature limit of the specific heat.


\end{document}